\RequirePackage{fix-cm}
\documentclass[smallextended]{svjour3}       
\smartqed  
\usepackage{graphicx}
\usepackage{natbib}

\begin{document}

\title{Optical Considerations for Large 3D Volumetric Particle Tracking Velocimetry}

\author{H. Abitan \and Y. Zhang \and S. l. Ribergård \and C. M. Velte}


\institute{H. Abitan \at
              first address \\
              Tel.: +45-23912203\\
              \email{haiab@mek.dtu.dk}  
}


\maketitle

\begin{abstract}
The continual increase in computational power and the improvement of algorithms for particle tracking in the past decade have been making it feasible to track larger amounts of particles in 3D Volumetric Particle Tracking Velocimetry (3D-PTV) experiments. Also, the relatively recent introduction of $15 \: \mu m$ Air Filled Soap Bubbles (AFSB) has been facilitating the usage of higher particle densities and hence the improvement of the spatial resolution of such measurements, compared with experiments that use $300 \: \mu m $ Hellium-Filled Air Bubbles (HFSB). The trend to conduct 3D-PTV experiments with ever increasing larger volumes or at higher particle densities with smaller particles sets an ever increasing strain on the power of the illumination source and upon the image analysis. On one hand it requires a reliable model to estimate the signal level that is measured on a CMOS detector from a Mie scattering particle. On the other hand it requires also a model for estimating the limiting factors upon the image resolution where a large amount of particles within a volume are mapped into a 2D image. Here, we present a model that estimates numerically the signal level on a CMOS detector from a Mie scattering particle within an arbitrary large volume in a 3D-PTV experiment. The model considers the effect of the depth of field, particle density, Mie factor, laser pulse energy and other optical parameters. Thereafter, we investigate the physical limit of the image resolution depending on the depth of field and the density of point-like particles. Finally, we supply three real lab examples that illustrate how to use the relevant expressions of the  models in order to estimate the signal level and the image resolution.

\end{abstract}

\section{Introduction}
\label{intro}
The continual increase in computational power according to Moor's law and the relatively recent development of the efficient algorithm, known as  Shake the Box by \citet{Schanz2016}, facilitates tracking of an increasing larger number of particles in 3D Volumetric Particle Tracking Velocimetry (3D-PTV). Since the particles number increases linearly with the volume of interest, 3D-PTV experiments can be done in ever increasing volumes. Also, the relatively recent introduction of $15 \: \mu m $ Air Filled Soap Bubbles (AFSB) by \citet{Barros2021} allowed to increase the particle density and improve the spatial resolution of such measurements \citep{Zhang2021} compared with $300 \: \mu m $ Hellium-Filled Air Bubbles (HFSB). The trend to conduct 3D-PTV experiments with ever increasing larger volumes or with smaller particles at higher densities, results in a meaningful decrease in the image signal of each particle and in a physical limit on the image resolution.
 
 The decrease in the signal level in large volumetric 3D-PTV has four main reasons: First, since the volume of interest in 3D-PTV needs to be illuminated with optical density that is similar to the optical density used in 2D-PIV experiments (with the same particles), the illumination of a volume requires pulses with energy that increases linearly with respect to the cross-section area of the volume of interest. A numerical example is useful to animate how the volume of interest relates to the required laser power.  Imagine a 2D-PIV experiment with a collimated laser beam (known as a laser sheet). The laser sheet has an elliptic cross-section area  $A=\pi \omega_a \omega_ b$, where $\omega_a$ and $\omega_b$ are the minor and major axes of the ellipse, respectively. The collimated  laser sheet can be viewed approximately as a thin rectangle cuboid with a small facet of area $A$ of width $\omega_a$ and height $\omega_b$ (a laser sheet). In 2D-PIV experiments $\omega_a=0.1 \: cm$ and $\omega_b=10 \: cm$ are typical values. If we would like to conduct the same experiment (with the same repetition rate), but in a rectangle cuboid volume with $\omega_a=10 \: cm$ and  $\omega_b=10 \: cm$, we would need a laser source with a pulse energy and average energy 100 times larger than the laser source that was used for 2D-PIV. 
 
  The second reason for the drop in the signal power relates to the particle size. Mie scattering depends on the second power of the diameter of the scattering particle. For example, when one conducts experiments with $15 \: \mu m $ AFSB in order to improve the spatial resolution of measurement of the flow field, the power scattered from a $15 \: \mu m $ AFSB particle falls by a factor of 400 compared with $300 \: \mu m $ HFSB particle (note the Mie Scattering converges to geometric optics for large particles). 

The third reason, relates to the imaging of particles within a volume. This requires the application of an aperture with a diameter $D_a$ smaller than the diameter of the lens of the camera. This is done in order to obtain a depth of field that contains the volume of interest. It means that less light can be gathered through the aperture from each particle by a factor $\left(\frac{D_a}{D_f}\right)^2$ where $D_f$ is the diameter of the lens.

The fourth reason that causes the image signal of a particle to drop relates to the depth of field. Particles that are at the edges of the depth of field appear as an enlarged and smeared spheres at the image plane. The optical power-density of the image of these particles is reduced. Hence, the optical power that falls into one pixel due to the smeared image of one particle drops. In particular, the particles that are at the far edge of the depth of field, will have the faintest signal. This is because the power density of their image will be reduced and because the cone of light that its apex is the particle and its base is the lens-aperture, will have the smallest solid angle.

 The image resolution (according to Rayleigh criterion) in volumetric 3D-PTV also decreases with the volume (more specifically, with the depth of field of the volume of interest). In 3D-PTV measurements, particles within a 3D volume are mapped into a 2D image plane. The density of the images of the particles in the image plane is clearly larger than the one obtained for 2D-PTV measurements with the same particles and particle density. The mere large number of imaged particles limits the image resolution. In addition, since one uses a camera with an aperture diameter that is adjusted for the required depth of field, the image of particles that are at the edges of the depth of field appears smeared and enlarged at the image plane (the surface of the CMOS detector). This reduces further the image resolution of the particles. The trend to develop 3D-PTV in larger volumes with smaller particles at higher densities means that the highest possible density will be for point-like light-source particles. This means that the limit on the particles density in such experiments would relate to the Rayleigh criterion, the depth of field and the Circle of Confusion ($CoC$) of point-like light-sources.

Here, we develop a quantitative model that estimates the signal level that is generated on a pixel of a CMOS detector by a scattering Mie particle located within a volume of interest. We first develop the model for 2D-PTV (a laser sheet) and then adapt it to 3D-PTV in a volume of interest. The model is particularly useful for finding the required pulse energy of the laser source in 2D and 3D PTV experiments. Next we analyse the image resolution of particles that flow in a volume of interest. We obtain an expression for the maximal particle density that depends on the depth of field of the volume of interest, the Circle of Confusion and the Rayleigh criterion. This expression is useful for understanding the optical factors that limit the spatial resolution in 3D-PTV measurements. Finally, we also supply three real lab examples that demonstrate the usage of the relevant expressions for predicting signal level and image resolution.

\section{Designing a 3D-PTV System}
\label{sec:1}
Designing a volumetric 3D-PTV flow measurement requires attention to four physical parameters of the phenomenon: the volume of interest $V$ (assume for clarity that it is a rectangular cuboid), the required spatial resolution $\Delta V$, the required temporal resolution of the measured velocity field $\Delta t$ and the maximum anticipated translational velocity of particles in the volume $u_{max}$. 

A 3D-PTV system is composed of four basic elements (Fig.1): 
\begin{itemize}
\item A pulsed laser that is characterized by a pulse energy $E_p$, pulse width $\tau$ and repetition rate $f_{rep}$. 
\item Optics to shape the laser beam and guide it into the volume of interest.   
\item Mie scattering particles with diameter $d$ and density $C$ (the particle density will set the spatial resolution $\Delta V.$). 
\item Two to four CMOS cameras. Each camera is characterized by detector dimensions (height $h$ and width $w$), responsivity $R(\lambda)$, frame rate, exposure time $\Delta t_{exp}$, pixel size $\Delta_{pix}$, read-out noise current $i_N$ and a lens with an effective focal length $f$ and transmission $T \left( \lambda \right)$.
\end{itemize}

In typical 3D-PTV experiments (Fig.1), a laser pulse is shaped by optical components into a beam with a cross-section that approximately matches the cross-section of the volume of interest. The shaped pulse is then guided into the volume of interest. The particles that are in the illuminated volume of interest scatter the laser light to all directions with an angular dependence according to Mie theory. A camera made of a lens, aperture and a CMOS detector points towards the volume of interest. Part of the light that is scattered from each particles manages to pass through the aperture of the camera. This light is imaged by the lens to generate images of each particle at the surface of its CMOS detector.
In order to estimate if we can get a detectable image of each particle within the volume of interest, we need to calculate the signal level in one pixel of the CMOS detector that is generated by any scattering particle within the volume. In particular, in 3D-PTV, we need to consider the particles that would have the faintest signal, i.e., particles at the far edge of the volume of interest. We also need to be able to distinguish between images of particles at the image plane. The Rayleigh criterion is often used to estimate the image resolution of particles that are in a plane. Here, we need to adapt the Rayleigh criterion for a resolvable image of particles that are in a volume.

\begin{figure*}
  \includegraphics[width=0.75\textwidth]{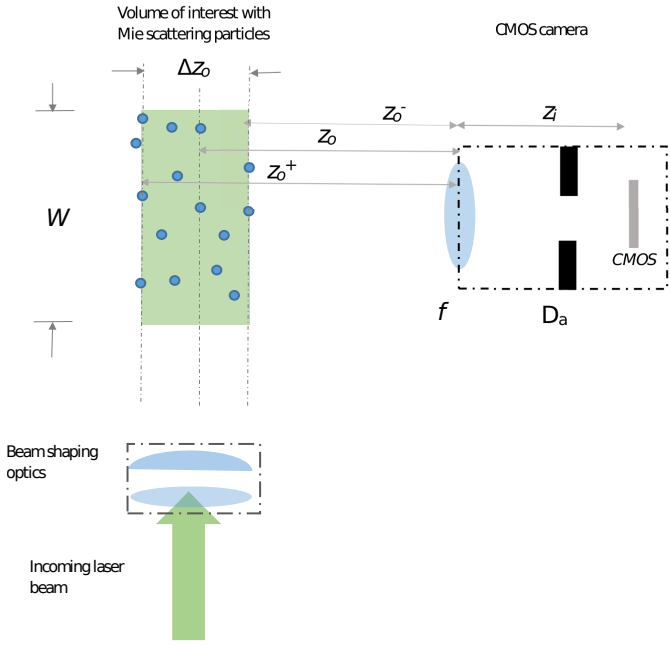}
\caption{A laser beam is shaped and guided into the volume of interest. The laser beam waist $\omega_a$ is assumed equal to the depth of field $\Delta z_o$. The working distance of the camera and its lens and the aperture are chosen so that the resulting depth of field $\Delta z_o$ equals $\omega_a$. The width of the field of view is $W$}
\label{fig:1}       
\end{figure*}


\subsection{Geometric Positioning of Camera}
\label{sec:2}
  In 2D-PTV it is a good practice to consider a Field of View (FoV) that has the same aspect ratio as of the CMOS detector.
Let $H$ and $W$ be the height and width of the FoV and $h$ and $w$ the height and width of the CMOS detector, respectively. Then ideally the magnification should be $M=\frac{h}{H}$. 
 A quick analysis of the geometry of the paraxial rays when imaging with a thin lens gives: 
\begin{equation}
\frac{H/2}{z_o}=\frac{h/2}{z_i}
\end{equation}
Which is the equation for magnification. Using this equation with the thin lens equation (see the appendix) gives that when we have a lens $f$, the required working distance is given by: 
\begin{equation}
 z_o=f\frac{M+1}{M}
\end{equation}

 In order to avoid lens aberrations in our image, the height of the field of view $H$ and the working distance $z_o$ needs to fulfil $\frac{H/2}{z_o}< 0.17$ . This ratio equals to the largest angle (in radians) of the paraxial ray (in the paraxial approximation $\tan \theta \approx \theta$ with less than one percent difference, as long as $\theta <0.17$). By inverting the last expression we get that as long as $z_o>\frac{H}{0.34}$, the image will have negligible lens aberrations. 
Often, CMOS cameras are used with a 'lens' that is made of several lenses, all together function as a single lens with shorter working distance (without image aberrations). Some lenses are marketed as a 'telephoto lens' and in principle consists of two groups of lenses (positive objective and negative eye piece) that together operate as a Galilean telescope. The telephoto lens reduces lens-aberrations and it facilitates flexible focusing. See more about it in the appendix. 
 
 In 3D-PTV we have to make sure that the whole volume will be observed without aberrations. Therefore, when the ratio of the depth of field to the working distance is non negligible, the FoV at $z_o^-$ should be used instead of the field of view at $z_o $.

Assume that our volume of interest $V$ is a rectangular cuboid with sides $a<b=c$, it is advantageous to point the collimated laser beam so that its elliptic cross-section with width $2\omega_a$ will equal the thin facet $a$ which in turn equal the depth of field $Delta Z_0$ as shown in Fig.1 (to optimize the delivery of optical power density). The camera should be aligned so that it will have the smallest possible depth of field (i.e. the depth of field  $Delta Z_0$ should equal $a$). In 3D-PTV one uses two or more cameras. The present analysis, as shown in Fig.1 is done for one camera positioned at right angle relatively to the laser beam. The considerations of this configuration should be adapted to a camera that has a different angle with respect to the laser beam.

\subsection{Selecting the Laser Source}
\label{sec:2}
Selecting a suitable pulsed laser source requires considering several features of the laser.  
\begin{itemize}
\item\textit{Repetition rate}: The laser repetition rate $f_{rep}$ and the frame rate of the CMOS camera should be equal. This frequency is set by the required time resolution of the experiment $\Delta t$  according to $f_{rep}=\frac{1}{\Delta t}$.  The CMOS exposure time $\Delta t_{exp}$ needs to be somewhere between the laser pulse duration $\tau$  and $\Delta t$.  . 

\item\textit{Laser pulse width} $\tau$: The maximal displacement of a particle during a laser pulse is when the maximal velocity $u_{max}$ is parallel to the object plane. In order to avoid streaks on the image from such a particle, we require that the maximal path length a particle will make during a light pulse, would be imaged into a length smaller than the dimension of half a pixel (at least). When this particle is moving in the object plane (2D-PTV) it means:

\begin{equation}
\tau u_{max}M < \frac{\Delta_{pix}}{2}
\end{equation}

This means that in 2D-PTV the pulse duration $\tau$ of the laser needs to fulfil:

\begin{equation}
\tau < \frac{\Delta_{pix}}{2Mu_{max}}
\end{equation}

In 3D-PTV, it is required to obtain an image signal from every particle within the volume of interest. When the depth of field is large, the $ z_o^- $ (as derived in the appendix) should be used for calculating the magnification of a distance at the plane $z_o^- $ where the magnification $M^-=\frac{z_i}{z_o^-}$ is the largest. Therefore, we require that the path done by particles with velocity $u_{max}$ at the plane $z_o^-$ would be smaller than half a pixel.

\begin{equation}
\tau u_{max}M^-<\frac{\Delta_{pix}}{2}
\end{equation}

This yields that the duration $\tau$ of the laser pulse must hold:

\begin{equation}
\tau<\frac{\Delta_{pix}}{2M^-u_{max}}
\end{equation}

\item\textit{The laser pulse energy} $E_p$:  the required pulse energy of the laser is calculated by finding what part of the pulse light is scattered by a Mie particle and what part of the scattered light reaches the CMOS detector to generate a signal current in one pixel. The image of one particle on the area of one pixel (or several pixels) is essentially an optical power. This optical power generates an electron current signal $i_S$ (the signal) in that pixel (or several pixels). The current-signal from one pixel equals the amount of absorbed optical power in that pixel times the responsivity of the CMOS detector at the wavelength of the laser $ i_S=P_{ab}R \left(\lambda\right)$ (the responsivity has the units of Ampere/Watts). 

Before calculating the signal current in one pixel of the CMOS detector as a function of the pulse energy of the laser, we note that on one hand, the Mie scattering factor depends on the intensity of the light $I$. On the other hand, the laser source is often specified by its pulse energy. We need to express the laser pulse energy in term of its optical intensity. We achieve that by two simplifying approximations. First, we approximate the shape of the laser pulse in the time domain to a square wave. Thus, the pulse energy and the pulse power of the laser are related by $E_p=P\tau$ (where the pulse power $P$ is now a constant during the pulse width). Second, we approximate the spatial Gaussian cross-section of the intensity of the laser beam to be a constant intensity but with an elliptic cross-section area with semi-major and semi- minor axes as of the actual Gaussian beam. Thus, the average intensity of such a beam is $I=\frac{P}{\pi \omega_a \omega_b}$. Where $P$ is the constant power and $\omega_a, \: \omega_b$ are the semi-minor and semi-major axes of the elliptical cross-section area of the original Gaussian laser beam. These two approximations give that the approximated optical intensity during a pulse length $\tau$ of the laser is given by: $I=\frac{E_p}{\pi \omega_a \omega_b \tau}$  

Now we are ready to track the path of the laser pulse from its output, until a signal is generated in a pixel of the CMOS detector due to scattering by a Mie particle.

First, the laser pulse is shaped (spatially) and then it is guided into the volume of interest. The shaping and guiding of the laser pulse will have optical losses $L$ due to reflection and scattering from surfaces of optical components.  Only $\left(1-L \right)$ of the original, averaged, pulse intensity $I=\frac{E_p}{\pi \omega_a \omega_b \tau}$ will reach the volume of interest. 
 
Second, the differential Mie scattering factor gives how much radiation will be scattered into a differential unit of a solid angle by one particle: 
 
\begin{equation}
 (1-L) \times I \times \frac{d^2}{\lambda^2} \frac{\partial \sigma}{\partial \Omega} 
\end{equation}

 where $(1-L)$ is the factor for the remaining intensity of the original pulse intensity $I$ of the laser, $d$ is the particle diameter, $\lambda$ is the wavelength of the laser, $\sigma$ is the scattering cross-section and $\partial \Omega$ is the differential solid angle. 
 
 When the solid angle $\Omega$ of the detector is large, one would need to integrate the differential Mie factor (Eq.7) with respect to the solid angle. However, in typical 3D-PTV, the solid angle extended from particles within the volume of interest toward the lens of the CMOS is very small. The particle at $z_o$ (the working distance in 2D-PTV) and the lens with diameter $D_f$ define a cone of light (the particle is at the apex and the lens area is the base of that cone). The solid angle of that cone is:
 
 \begin{equation}
  \Omega_{par}=\frac{\pi \left( D_f  \right)^2} {16 \pi z_o^2}
 \end{equation}

 It is typically very small (milli steradians). Therefore, we are justified to approximate the differential Mie scattering factor to be a constant in that small solid angle and avoid integration. We use the scattering coefficient at the angular direction of the Mie theory (90 degrees in Fig.1) and multiply it by the solid angle  $\Omega_{par}$ extended by the lens toward the particle.  
 
Third consideration concerns with optical losses due to the optics of the signal gathering. The ‘lens’ of a CMOS camera usually consists of a combination of several types of lenses (to compensate for lens aberrations). This results in a considerable transmission loss. Let us designate by $T \left( \lambda \right)$ the transmission of the camera's ‘lens’ at the laser wavelength $\lambda$.

Finally, one needs to consider the ratio of the area of one pixel to the area of the image of one particle, I.e., the Fill Factor Ratio (FFR). When the image area is larger than a pixel area, the amount of optical power that will be absorbed by one pixel will be at best proportional to the ratio of the area of one pixel $ \Delta_{pix}^2 $ to the area of a particle image $0.25 \pi M^2 d^2$. We designate this ratio by $FFR$.  

 According to the four aforementioned factors, the amount of optical power that arrives to one pixel from one scattering particle is given by:
\begin{equation}
P_{ab}=\left( 1-L \right)\times I \times \frac{d^2}{4\lambda^2} \frac{\partial \sigma \left( \theta \right)}{\partial \Omega} \times\Omega_{par}  \times T \left( \lambda \right) \times FFR 
\end{equation}

Substituting into Eq.9  the explicit expressions for the intensity of the laser and for the solid angle $\Omega_{par}$, then multiplying it by the responsivity, we obtain the signal current in one pixel for 2D-PTV experiment:

\begin{equation}
i_S=\left( 1-L \right)\times \frac{E_p}{\pi \omega_a \omega_b \tau} \times \frac{d^2}{4\lambda^2} \frac{\partial \sigma \left( \theta \right)}{\partial \Omega} \times   \frac{D_f^2} {16 z_o^2}  \times T \left( \lambda \right) \times FFR \times R \left( \lambda \right) 
\end{equation}

\end{itemize}

To increase the signal current $i_S$ in 2D-PTV measurement, one should aspire to minimize the optical loss due to beam shaping and guiding. One should aspire to work with a short pulse (at a given pulse energy), optimize the cross-section of the beam so it would be as thin as possible (i.e. a suitable waists of the laser beam), select a particle diameter and laser wavelength that maximize the function $\frac{d^2}{\lambda^2} R \left(\lambda \right) T \left(\lambda \right)FFR$, have a working distance $z_o$ as short as possible (yet avoid optical distortions at too short working distance) and increase lens diameter $D_f$ to maximize signal gathering. 
 
In digital image analysis SNR=10 (the Signal to Noise Ratio $SNR=\frac{i_S}{i_N•}$ ) is considered to be acceptable and SNR=40 is considered to be excellent. Hence, one should aim to have a SNR that lies between 10 to 40.


\section{The signal level in 3D PTV}
By applying two minor changes, Eq.10 can be modified for estimation of the signal current $i_S$ in 3D-PTV experiments: 
First, we want to assure a signal from all the particles within the volume of interest. The particles at the far edge of the depth of field will have the smallest solid angle $\Omega_{par}$ with respect to the diameter of the aperture of the lens. Therefore, the signal from particles at the far edge of the depth of field represents the signal level required for 3D-PTV. We employ this understanding by replacing $z_o$ with $z_o^+$  and by replacing the lens diameter $D_f$ with the aperture diameter $D_a$ in  Eq.10 (the explicit expression for $z_o^+$ is found from  Eq.20 and Eq.21 in the appendix). One should use it when the ratio of the depth of field to the working distance is not negligible.

Second, when the diameter of the aperture is decreased, particles that are not at the object plane will appear as smeared circles on the image plane. In particular, particles at the far edge of the depth of field (at $z_0^+$) will have a diameter given by $M^+d+CoC $. This image diameter is larger than the image diameter of particles at the object plane $z_0$. The fill factor ratio of the image of a particle at the edge will be smaller than the fill factor ratio for particles at the object plane. We mark this by replacing $FFR$ with $FFR\left( M^+d+CoC)\right)$ in Eq.10 and thus:

\begin{equation}
i_S=\left( 1-L \right) \times \frac{E_p}{\pi \omega_a \omega_b \tau} \times \frac{d^2}{4\lambda^2} \frac{\partial \sigma \left( \theta \right)}{\partial \Omega} \times   \frac{ D_a^2} {16 \left( z_o^+ \right)^2} \times T \left( \lambda \right) \times FFR\left( M^+d+CoC)\right) \times  R \left( \lambda \right) 
\end{equation}

In 3D-PTV measurements $N$ particles from a 3D volume are mapped onto a 2D image at the image plane. The thickness of the volume of interest is designated by $\Delta z_o$ (see Fig.1) and it equals to the depths of field of the imaging lens. The depth of field is determined by the diameter of the aperture of the lens $D_a$ and the acceptable Circle of Confusion of an ideal point-like light-source (see the appendix). Accordingly, particles at the object plane will have at the image plane a diameter of $Md$ and particles located at the far edges of depth of field will have a diameter $M^+d+CoC$ at the image plane. The diameter $CoC$ affects the signal level as was explained above. It also affects the image resolution. Since the image of the particle appears larger. Next, the interplay between the diameter $CoC$, depth of field and the image resolution will be quantified.

\section{Image Resolution and Particle Density in 2D and 3D-PTV}
We assume that in 3D-PTV experiments particles with an average diameter $d$ are distributed with an average density within a volume of interest. An image of these particles is a mapping of particles from a 3D volume onto a 2D surface (the CMOS detector). We ask what is the maximal particle density  that will give us a 2D image with particle images that are resolvable (the particles are considered as point sources).

The particle density $C$ and the average distance between the particles $D_R$ are related by $C=\frac{1}{D_R^3}$.  Therefore, if we can find the minimal average distance between particles where their image is just resolvable (according to Rayleigh criterion), we can find the maximal possible density and the limit of the spatial resolution $\Delta V $ of the flow measurement. The maximal particle density is achieved when the particles are considered as point-like light-sources. Then, the two limiting factors are the Rayleigh resolution criterion (see \cite{Pedrotti1993}) and the interplay between the Circle of Confusion and the depth of field. 

Rayleigh criterion for angular resolution $\Delta \theta_R$ is an ad hoc criterion, yet, empirically proven (see \cite{Pedrotti1993}). The criterion tells us that when two adjacent point-like light-sources are imaged onto a screen, one will observe a diffraction pattern made of two light discs (called Airy disks), each surrounded by dark and lighted rings. When the maximum of the first Airy disc falls on the first dark ring of the second Airy disc, one can just barely deduce that the diffraction pattern is due to two different point sources. In that case, the angular distance between the two light sources is found to be: 

\begin{equation}
\Delta \theta_R=1.22 \frac{\lambda}{D_a}
\end{equation}

We examine two particles that lie in a thin laser sheet that coincides with the object plane at $z_o$ (as in 2D-PTV experiments). The corresponding Rayleigh distance between these two particles in the object plane is $ D_R^o=\Delta \theta_R z_o$ and the distance between the two Airy discs in the image plane is $ MD_R^o$. In order to distinguish between two Airy discs that are $M D_R^o$ apart on a CMOS surface, the distance between the two discs must be at least one pixel. This has two implications: first, the  pixel size must be smaller than $M D_R^o$. Second, if the thickness of the laser sheet is $D_R^o$, the maximal particle density in the volume of the laser sheet can not exceed $C_{2D-PTV}=\frac{1}{\left( D_R^o \right)^3}$. In this density, it is unlikely that images of particles within this laser sheet will merge and convolve on the image plane. Hence, the maximal particle density for such a laser sheet is $C_{2D-PTV}$. The corresponding maximal number of particles that can be imaged is $N_{2D-PTV}=C_{2D-PTV}WHD_R^o$. In 3D-PTV the depth of field $\Delta z_o$ is usually much larger than $D_R^o$. 
If we want to have an image of all particles within that volume, where their images do not convolve on the image plane, the number of particles in the volume can not exceed $N_{2D-PTV}$. This leads us to conclude the maximal density in 3D-PTV measurement with depth of field $\Delta z_o$ can not exceed 

\begin{equation}
C_{3D-PTV}=\frac{C_{2D-PTV}}{\Delta z_o}D_R^o
\end{equation}

Two remarks are worthwhile. First, the condition about the thickness of the laser sheet (in 2D-PTV experiment ) introduces a new limitation on the maximal width $W$ of the FoV of the thin volume of interest (in 2D-PTV experiment). The thickness of a Gaussian laser beam within a range centred about a beam waist $\omega_0$ is considered nearly constant in a range known as the Rayleigh range \citep{Hawkes1995}. This gives that the maximal width of the field of view $W$ where $2 \omega_a < D_R$ equals to the Rayleigh range, hence: $W<\frac{1}{4} \frac{\pi \left(D_R^o \right)^2}{\lambda}$.  

Second, we note that in 3D-PTV two particles that are located at the $z_o^+$ plane and with angular distance $\Delta\theta_R$ will have the distance $ D_R^+=\Delta\theta_R z_o^+$, hence $D_R^+>D_R^o$. Also, their diameter is increased to $CoC$, due to the depth of field. This means that they will not satisfy Rayleigh criterion. To ensure that particles in the object plane will satisfy Rayleigh criterion and Particles at the far edge of the depth of field will also be resolveable, we realize that the distance between the two Airy discs of particles at the object plane needs to be increased by one circle of confusion. The corresponding angular distance is:

\begin{equation}
\Delta\theta_{R-CoC}^+=\frac{\Delta \theta_R f+CoC}{f}
\end{equation}

We use $\Delta\theta_{R-CoC}^+$  to calculate that the distance between two particles at the $z_o^+$ plane needs to be

\begin{equation}
 D_{R-CoC}^+=\Delta\theta_{R-CoC}^+z_o^+
\end{equation}

This means that our estimation for the maximal density in 3D-PTV (Eq.13) should be used with $D_{R-CoC}^+$.

\section{Conclusion and outlook}
We developed a practical model to calculate the signal level on a CMOS detector in 2D-PTV experiment with a laser as the illuminating source. We extended that model to be used in 3D-PTV experiments with an arbitrary depth of field. In 3D-PTV, one needs to consider the signal level from particles with the faintest signal, i.e. particles at the far edge of the depth of field.  We also modelled the image resolution in both 2D and 3D-PTV experiments (for point-like light-source particles). Also here, in 3D-PTV one needs to consider the image resolution due to particles at the far edge of the depth of field. We found that both signal level and image resolution are compromised due to the circle of confusion. Never the less, we obtained a expressions to estimate the signal level and image resolution of particles within an arbitrary depth of field. 
The spatial resolution of 2D-PTV and 3D-PTV measurements depends on the the maximal possible particle density which in turn depends on the image resolution. We obtained an expression to estimate the maximal particle density, depending on image resolution for both 2D and 3D-PTV measurements. 

The presented models are simplified since they consider a Gaussian laser source and point like particles. In the future we intend to extend the present model and consider Light Emitting Diodes (LEDs) as the illumination source and particles of any diameter. LEDs have a wide spectral distribution and a Lambertian beam shape. Also, LEDs are often used as LEDs array for large volume 3D-PTV experiments. Therefore, a model for the signal level and image resolution of particles within a volume that are illuminated by LEDs will require additional elaborations to the present models.   

\begin{acknowledgements}
H. Abitan acknowledges financial support from the Poul Due Jensen Foundation: Financial support from the Poul Due Jensen Foundation (Grundfos Foundation) for this research is gratefully acknowledged.
Y.Zhang, S. l. Ribergård and C. M. Velte acknowledge financial support from the European Research council: This project has received funding from the European Research Council (ERC) under the European Unions Horizon 2020 research and innovation program (grant agreement No 803419).

The authors would like to thank Benny Edelstein for meticulous proof reading of the text and math. 
\end{acknowledgements}

\section{Appendix: Thin lens, Depth of Focus, Depth of Field and CoC}
\subsection{Thin Lens}

The thin lens equation is essential for understanding the image generation of the Mie scattering particles in volumetric PTV.  Fig.2 illustrates the main concepts involved with a thin lens.

Three planes are defines in perpendicular to the optical axis. The plane where the object lies is called the Object Plane; and similarly, the Lens Plane and the Image Plane.
According to geometric optics, objects consists of point-like light-sources that lie in the Object plane. Each point source of the object emits light rays to all direction. The rays of each point source which define a cone where its apex is the point-like light-source and its base is the area of the lens are of specific interest. Ideally, all the rays within that cone will be imaged into a point in the image plane.
 A thin lens has two spherical boundaries. Light rays that are incident on the surface of a thin lens will be refracted and change direction as they are transmitted through the lens. 
 \begin{figure}
\includegraphics{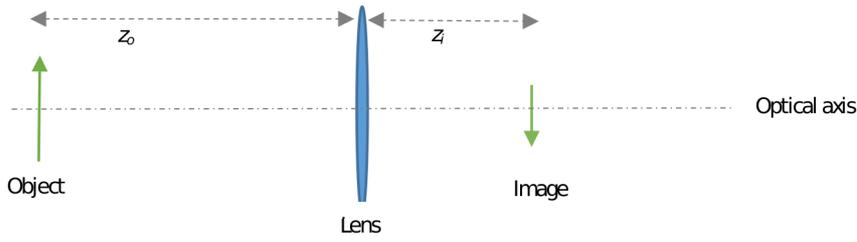}
\caption{A positive thin lens. The object is at $z_o$ and its inverted image is at $z_i$ }
\label{fig:2}       
\end{figure}
 According to the thin lens model, the light rays from one point source at distance $z_o$ will be refracted at the lens and focused into a point at the other side of the lens at distance $z_i$. The thin lens is characterized by a focal length $f$  where $z_o,z_i$ and $f$ hold, in first approximation, the thin lens equation:
 
\begin{equation}
\frac{1}{z_o}+\frac{1}{z_i}=\frac{1}{f}
\end{equation}

A digital camera consists of a CMOS detector, a lens and an aperture. When we point the camera toward an object at distance $z_o$ away from the lens. In order to view the object ‘in focus’ we need to place the CMOS detector a distance $z_i$ away from the lens at the opposite side of the lens (according to the thin lens equation). Then all point sources of the object will be mapped into image points in the image plane. The image will be said to be 'in focus'. 

Fig.3 shows the apex rays from three different point-like light-sources that are on the optical axis. Only the  rays that can just go through the aperture are shown. Each point source has a conjugated image point on the other side of the lens. The locations of the point sources and their conjugated image points obey the thin lens equation.

\begin{figure}
\includegraphics{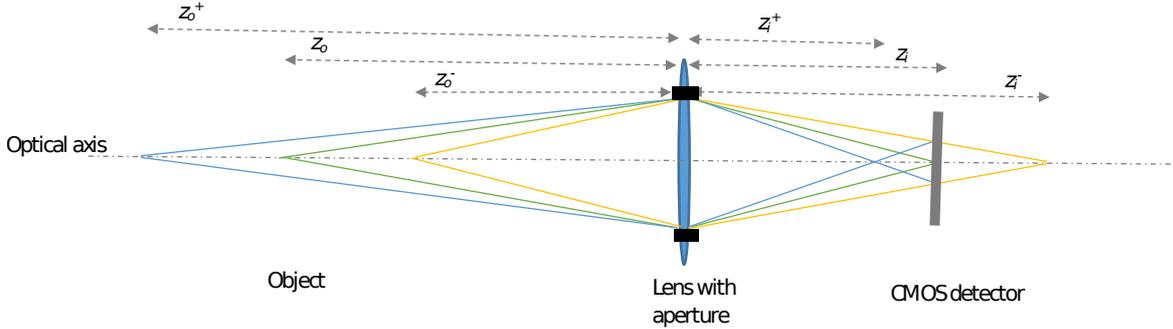}
\caption{The apex rays of three point sources (on the left hand side of the lens) are mapped into three image points. A screen is put at the distance $z_i$ where the point source $z_o$ is focusing and the other two source points generate a blurred circle of light called the circle of confusion}
\label{fig:3}       
\end{figure}

\begin{equation}
\frac{1}{z_o}+\frac{1}{z_i}=\frac{1}{f}, \:  \frac{1}{z_o^+}+\frac{1}{z_i^+}=\frac{1}{f}, \: \frac{1}{z_o^-}+\frac{1}{z_i^-}=\frac{1}{f}
\end{equation}

The point-like light-source of the object at $z_o$ is imaged into a point-like light-source at $z_i$ on the CMOS detector (the green rays). However, the other two point-like light-sources are imaged into blurred circles at the image plane $z_i$, known as the circle of confusion with diameter $CoC$.
The blue rays on the right hand side cross each other and define two similar isosceles triangles. One has the $CoC$ as the base and the other has the aperture diameter $D_a$ as the base. The following relation is easily deduced from the geometry:

\begin{equation}
\frac{CoC}{z_i -z_i^+}=\frac{D_a}{z_i^+}
\end{equation}

The yellow rays also define two isosceles similar triangles. The small one has the CoC as its base. The larger one has $D_a$  as its base. We can find the following relation from their similarity:

\begin{equation}
\frac{CoC}{z_i^- -z_i}=\frac{D_a}{z_i^-}
\end{equation}

Eq.18 and Eq.19 give: $z_i^+=\frac{D_a}{D_a+CoC}z_i$ and $z_i^-=\frac{D_a}{D_a-CoC}z_i$ .

The thin lens equation for $z_o,z_i$ yields $z_i=\frac{fz_o}{z_o-f}$. We plug it in the expressions for $z_i^+$ and$z_i^-$ and obtain
\begin{equation}
z_i^+=\frac{D_a}{\left(D_a+CoC \right)} \frac{fz_o}{\left(z_o-f \right)} \: , \: z_i^-=\frac{D_a}{\left(D_a-CoC \right)}\frac{fz_o}{\left(z_o-f \right)}
\end{equation}

 Their difference $\Delta z_i=z_i^- - z_i^+$  gives the depth of focus. I.e., a segment along the optical axis where a point source from the object plane is imaged into a circle that is smaller than the circle of confusion. $\Delta z_i$ is useful when one tilts the detector with respect to the optical axis, as in Schinflug technique.

The depth of field $\Delta z_o=z_o^+ - z_o^-$ is a segment along the optical axis where the object lies so that any point source within the depth of field will appear in the image plane as a circle of light with a diameter smaller than $CoC$. We use Eq.17 to express $z_o^+$ as a function of $z_i^+$ and $f$. Similarly we express $z_o^-$ as a function of $z_i^-$ and $f$:

\begin{equation}
 z_o^+=\frac{fz_i^+}{z_i^+ -f}, \: \: z_o^-=\frac{fz_i^-}{z_i^- -f}
\end{equation}

 We plug into Eq.21 the expressions for $z_i^+$ and  $z_i^-i$  (Eq.20) and obtain the expressions for $z_o^+$ and $z_o^-$ as functions of the   aperture diameter $D_a$, diameter of the circle of confusion $CoC$, focal lens $f$ and the working distance $z_o$. finally we use $\Delta z_o=z_o^+ - z_o^-$ to calculate the depth of field.   

Sometimes it is more practical to invert Eq.21 and express $CoC$ as a function of the depth of field, working distance, aperture diameter and focal length.

\begin{equation}
CoC=-\frac{D_a z_i}{\Delta z_o}+\sqrt{\left( \frac{D_a z_i}{\Delta z_o} \right)^2+\left(\frac{D_a z_i}{f}-D_a \right)^2}
\end{equation} 

where $z_i=\frac{fz_o}{z_o-f}$

\subsection{Telephoto Lens}
Sometimes one needs a lens with large focal length in order to view a distant field of view with relatively large image.  This means that the image plane will be also at $z_i \approx f$, which would be also large. It is not practical to place the CMOS censor so far from the lens of the camera. Hence, one use a Galilean telescope which is made of two lenses: one positive $f_1$ and one negative $f_2$ that are a distance $l $ apart. Ray tracing analysis \cite{Pedrotti1993} of such two lenses shows that the Back Effective Focal Lens ($BEFL$) of the two lenses is:

\begin{equation}
\frac{f_1 f_2}{l-\left(f_1-f_2 \right)}
\end{equation}

when the lenses are put $\left(f_1-f_2 \right)$ apart, the $BEFL$ is at infinity. However, by tuning the distance between the two lenses a bit away from that point, one reaches a point where the $BEFL$ is comfortably tuned by small increase or decrease of $l$.  A Telephoto lens is in principle a Galilean telescope. It is made of two groups of lens: one positive and one negative. Other lenses are added in order to compensate for lens aberrations (spherical, chromatic and coma). But the basic optics of the Telephoto lens is as of a Galilean telescope. The important point to remember is that their $BEFL$ is controlled in order to put in focus objects at different working distance $z_0$.     
  
%

%

\subsection{Numerical Example of 2D-PTV}
\label{numerical_example}

\begin{center}
\begin{tabular}{ |c|c|c| } 
 \hline
\textbf{ Volume of Interest}  &  &  \\ 
 volume & $a=0.1 \:cm \: b=c=10 \: cm$ & rectangular cuboid laser sheet\\ 
 Max velocity  &$u_{max}=10 \frac{m}{s}$  &  \\ 
 time resolution $\Delta t=0.5 \: ms$ &  &  \\ 
 spatial resolution  & $\Delta V=?$ &  \\ 
 \hline
 \textbf{Laser}  &  & \\
 wavelength & $532 \: nm$ & green\\
 laser waist & $\omega_a=0.1, \omega_b=10 \:cm $ & elliptic \\
 repetition rate  &$f_{rep}=?$  & \\
 pulse duration & $\tau=?$ & \\
 pulse energy  & $E_p=?$ & \\

 \hline
 \textbf{Particles} & & \\
 average diameter & $d=1 \mu m$ & \\
 material & DEHS & \\
 refractive index & 1.45 at $532 \: nm$& \\

\hline
 \textbf{CMOS Camera} &  & \\
 dimensions &  $20.48 \times 20.48 \: mm$ & \\
 pixels No. & $1024 \times 1024 $ & $9000 fps$\\
 pixel size & $\Delta_{pix} = 20 \: \mu m$ & \\
 Responsivity & $R \left( \lambda =532 \: nm \right) = 0.78$ & \\

\hline
 \textbf{Optics} &  & \\
 beam guiding loss &$L=0.2$  & \\
 signal transmission & $T \left( \lambda =532 \: nm \right) = 0.8$.  & \\
\hline
\end{tabular}
\end{center}

First, we find the minimal working distance, where there wont be lens aberrations, by using $z_o>\frac{H}{0.34}$. This gives: $z_o >294 \:mm$. 
The Field of View (FoV) is $10 \times 10 \: cm^2 $. It has the same aspect ratio as of the CMOS area. 
In order to use the whole area of the CMOS detector, we would like to work with magnification $M=$(Width of CMOS)/(Width of the FoV)$= 0.2$. 
Using Eq.2 we find that $z_o=6f $, accordingly. If the lens we have is $f=60 \: mm$ then $z_o=360mm$ and our image at $z_i=72 \: mm$ will be free of aberrations.
 
\textit{The Laser Repetition Rate}: $f_{rep}=\frac{1}{0.5 \cdot 10^-3}=2kHz$.

\textit{The Pulse Duration}: $\tau$ is found from Eq.4: $\tau< \frac{20 \times 10^{-6}}{2 \cdot 0.2 \cdot 10}=0.25 \mu s$. We choose a laser with $\tau=0.1 \: \mu s$. 

\textit{The required laser pulse energy}: 
We assume that the Gaussian laser sheet with $\omega_a=0.1 \:cm$ and $\omega_b=10 \:cm$ overlaps with the rectangular cuboid $V$ almost in an ideal level (i.e. we approximate). 
The differential Mie scattering factor at $90 \deg$ is calculated from the free-ware of Philip Laven program (www.philiplaven.com). The program gives a Mie factor $5 \times 10^{-18} \frac{W}{cm^2}$  at 1 meter.  We use a lens with a diameter $D_f=50 mm$ and working distance $z_o=360 \: mm$ to calculate that the solid angle is $\Omega_{par}=1.2 \cdot 10^{-3} \: str$. The $FFR=1$ (since the area of the image of one particle falls into one pixel). Plugging these numbers and those from table 1 into Eq.10: 
$i_S=0.8 \times \frac{E_p}{3.14 \cdot 10^{-7} }\times  5 \cdot10^{-18} \times 3.6 \cdot 10^{-3} \times 0.85 \times 1 \times 0.78$ or $i_S=3 \times 10^{-14} \times E_p$ Amperes. The electron count from a pixel for $E_p=1 \: mJ$ is found by dividing $i_S$ with the electron charge $e=1.6 \times 10^{-19}$. We obtain that we will have $187$ electrons counts from an image of one particle. The Photron Nova S9 has an electron noise-count of about 20  per pixel. Hence we will have $SNR=9$ at $1 \: mJ$ laser pulse energy. 

\textit{Image Resolution and Density}: the minimal distance of the particles at the Rayleigh limit is $ D_R^o=1.22\frac{0.532 \cdot 10^-6}{0.035} \times 0.36=4.5 \: \mu m $. This correspond to a distance $M D_R^o=0.2 \times 4.5=0.72 \: \mu m$ in the image plane. However, we note that the pitch of our CMOS sensor is $20 \:  \mu m$. Hence due to our CMOS resolution we would be able to distinguish only between two particles with a distance larger than $ \Delta_{pix}/M=20/0.2=100 \: \mu m$.  Also, we note that the thickness of our laser sheet is $1 \:mm$. It is $10$ times larger than the pixel resolution. If we will work with particle density that correspond to $D_R^o=100\: \mu m$, most likely that the image of particles within the $1 \:mm$ thick laser sheet will convolve. If the diameter of the elliptic laser beam at its waist will be reduced to the minimal distance between particles  ($2\omega_a=D_R^o=100 \: \mu m$), then the images of particles from the thin Rayleigh Range of width $W<\frac{1}{4} \frac{\pi D_R^2}{\lambda}=14 \:  mm$, most likely will not convolve in the image plane.

\subsection{Numerical Example of a Small Volume 3D-PTV}
\label{numerical_example}

\begin{center}
\begin{tabular}{ |c|c|c| } 
 \hline
\textbf{ Volume of Interest}  &  &  \\ 
 volume & $a=0.3 \:cm  \: b=c=2.7 \: cm$ & rectangular cuboid \\ 
 maximum velocity  &$u_{max}=3 \frac{m}{s}$  &  \\ 
 time resolution $\Delta t$ & $0.00025 \: s$ &  \\ 
 spatial resolution  & $\Delta V=?$ &  \\ 
 \hline
 \textbf{Laser}  &  & \\
 wavelength & $532 \: nm$ & green\\
 laser waist & $\omega_a=0.15 \: cm $ & circular \\
 repetition rate  &$f_{rep}=?$  & \\
 pulse duration & $\tau=?$ & \\
 pulse energy  & $E_p=?$ & \\

 \hline
 \textbf{Particles} & & \\
 average diameter & $d=1 \mu m$ & \\
 material & DEHS & \\
 refractive index & 1.45 at $532 nm$& \\

\hline
 \textbf{CMOS Camera} &  & \\
 dimensions &  $27.6 \times 26.3 \: mm$ & \\
 pixel No.  & $2048 \times 1952 $ & $ 4,855fps$\\
 pixel size & $\Delta_{pix} = 13.5 \: \mu m$ & \\
 responsivity & $R \left( \lambda =532 \: nm \right) = 0.23$ & \\

\hline
 \textbf{Optics} &  & \\
 beam guiding loss &$L=0.04$  & \\
 signal transmission & $T \left( \lambda =532 \: nm \right) = 0.8$  & \\
 objective lens diameter &$D_f=70 \: mm$ & \\
\hline
\end{tabular}
\end{center}

First, we need to find the minimal working distance where there wont be lens aberrations. By using $z_o>\frac{H}{0.34}$ we find $z_o >80 \:mm$. Due to experimental constrains we need to work from a distance of $1000 \:mm$. Our volume of interest has a Field of View (FoV) $2.7 \times 2.7 \: cm^2 $ that essentially equals the CMOS area. Hence, the magnification $M=1$. Using Eq.2 we find that for $z_0=1000 \:mm $, the required lens would be $f=500 \:mm$. Using the thin lens equation we find that $z_i=1000 \: mm$ which is very large to be used with a CMOS sensor (the sensor will need to be 1 meter away from the lens). The solution is to use a Galilean telescope. The Nikkor 300 mm Telelphoto lens is basically an elaborated Galilean telescope (see a short explanation about such a telescopic lens in the appendix). It has a Front Effective Lens $FEFL=300 \: mm$. When it is continued with Nikkor teleconverter 2M lens, it's $FEFL=600 \: mm$. By turning the focusing cylinder knob, we tune the $BEFL$ so that we would see the field of view in focus at a distance of 240 mm from the edge of the Nikkor teleconverter 2M lens.

\textit{The Required Laser Repetition Rate:}  $f_{rep}=\frac{1}{0.00025}=4kHz$.

\textit{The Required Pulse Duration:} $\tau$ is found from Eq.6: $\tau< \frac{13.5 \times 10^{-6}}{2}=6.75 \: \mu s$. We choose a laser with $\tau=0.01 \: \mu s$.

\textit{The Required Laser Pulse Energy:} We assume a circular Gaussian laser beam with $\omega_a=\omega_b=0.15 \:cm$ that approximately fills the volume of interest $V$. In this example the ratio of the depth of field to the working distance is very small. Hence we can use $z_o$ instead of $z_o+$ in Eq.11.  
The differential Mie scattering factor at $ang{90}$ is calculated by using Philip Laven free-ware programs (www.philiplaven.com). The program  gives a Mie factor $5 \cdot 10^{-18} \frac{W}{cm^2}$ at 1 meter. The f-number used is $5.6$, hence the aperture is about $D_a=\frac{300}{5.6}=53 \: mm$ and $\Omega_{par}=3.4 \cdot 10^{-4}$. The $FFR=1$. Plugging these numbers and those from table 2 into Eq.11, we obtain: 

$i_S=0.96 \times \frac{E_p}{7 \cdot 10^-10} \times  5 \cdot10^{-18} \times 1.7 \cdot 10^{-4} \times 0.8 \times 1 \times 0.23$

or $i_S=7 \times 10^{-13} \times E_p$ Amperes. The electron count from a pixel for $E_p=1 \: mJ$ is found by dividing $i_S$ by the electron charge $e=1.6 \times 10^{-19}$. We obtain that we will have 437 electrons counts from an image of one particle per $1 \: mJ$ pulse energy.(the image falls into one pixel). The  electron  noise-count in this camera is 29.7 per pixel. Hence,$SNR=14$ and we will have an excellent PTV signal.

\textit{Optical resolution and density:}
First we use Eq.22 to find that for $z_o=1000 \: mm, \: D_a=5.3 \: mm \: , f=300 \: mm$ and depth of field $\Delta z_o=3 \: mm$ the circle of confusion is $CoC=3.4 \: \mu m $.  We further find from Eq.21 that  $z_o^-=998.5$ and $z_o^+=1001.5 \: mm$.
 The Rayleigh distance at the object plane $z_o$ is $ D_R^o=1.22\frac{0.532 \cdot 10^-6}{0.0053} \times 1000= 12.2\mu m $. Because $M=1$, the images of the two particles (Rayleigh discs) would have the same distance in the image plane. Our sensor has a pitch of  $13.5 \: \mu m$. The pixel resolution nearly matches the Rayleigh resolution. 
The images of point sources from $z_o^-$ and $z_o^+$ will have a diameter of $3.4 \: \mu m $ (circle of confusion). Using Eq.14 we find that due to $CoC=3.4 \: \mu m $ the minimal Rayleigh distance at $z_o^+$ plane is $D_{R-CoC}^+=12.2+3.4=15.6 \: \mu m $. 
We use it to find that the maximal particle density for laser sheet with thickness $D_{R-CoC}^+$ is  $C_{2D-PTV}=\frac{1}{\left( D_{R-CoC}^+\right)^3}$.  
The maximal density in cubic $cm$ is obtained by using Eq.13 with $D_R^o$ replaced by $D_{R-CoC}^+$:

$C_{3D-PTV}=\frac{1}{\left( D_{R-CoC}^+ \right)^2 \Delta z_o}=\frac{1}{\left(15.6 \cdot 10^-4\right)^2 \times 0.3}=4.4 \cdot 10^5 \frac{1}{cm^3}$

\subsection{Numerical Example of a Large Volume 3D-PTV}
\label{numerical_example}

\begin{center}
\begin{tabular}{ |c|c|c| } 
 \hline
\textbf{ Volume of Interest}  &  &  \\ 
 volume & $a=10 \:cm \:  b=c=30 \:cm$ & rectangular cuboid \\ 
 velocity  &$u_{max}=30 \frac{m}{s}$  &  \\ 
 time resolution $\Delta t$ & $0.00025 \: s$ &  \\ 
 spatial resolution  & $\Delta V=?$ &  \\ 
 \hline
 \textbf{Laser}  &  & \\
 wavelength & $532 \: nm$ & green\\
 laser waist & $\omega_a=5 \: cm \: \omega_b=15 \: cm$ & elliptic \\
 repetition rate  &$f_{rep}=?$  & \\
 pulse duration & $\tau=?$ & \\
 pulse energy  & $E_p=?$ & \\

 \hline
 \textbf{Particles} & & \\
 average diameter & $d=15 \mu m$ & \\
 material & Air Filled Soap Bubbles & \\
 refractive index & 1.33 at $532 nm$& \\

\hline
 \textbf{CMOS Camera} &  & \\
 dimensions &  $27.6 \times 26.3 \: mm$ & \\
 pixel No.  & $2048 \times 1952 $ & $ 4,855fps$\\
 pixel size & $\Delta_{pix} = 13.5 \: \mu m$ & \\
 responsivity & $R \left( \lambda =532 \: nm \right) = 0.23$ & \\

\hline
 \textbf{Optics} &  & \\
 beam guiding loss &$L=0.04$  & \\
 signal transmission & $T \left( \lambda =532 \: nm \right) = 0.8$.  & \\
\hline
\end{tabular}
\end{center}

First, by using $z_o>\frac{H}{0.34}$ we find that the working distance $z_o$ should be larger than $882 \:mm$ and it corresponds to an angle of view of $19 \deg$. 
However, the large volume of interest in this experiment is about 100 times thicker than a typical 2D-PTV volume. It means that we would need a laser source with a pulse energy and average energy that are 100 times larger, compared with a similar 2D-PTV experiment. Since, according to Eq.11 the signal level depends inversely on the second power of $z_o^+$. We would like to work with $z_o$ as small as possible. I.e. we would like to place the camera as close as possible to the volume of interest. This will increase the signal level. We found that a Nikkor 35 mm lens (from Nikon Corporation) can have aberration free image down to a distance of 300 mm (this lens is made of 8 lenses that together decrease lens aberrations and facilitate an angle of view of $ang{42}$).

The dimensions of our CMOS sensor are $2.7 \times 2.7 \: cm^2 $ and FoV dimensions are $30 \times 30 \: cm^2 $. Thus the magnification is $M=$(Width of CMOS)/(Width of the FoV)$=0.09$. Using Eq.2 we find that for $f=35 \:mm$, the required working distance is $z_o=424 \: mm$.
When the $f/N=11$, the aperture diameter is 3.18 mm. We plot Eq.21 as shown in Fig.4 for $z_o=424 \:mm, f=35 \: mm$ and $D_a=3.18 \:mm$. We find from the plot that for depth of field $\Delta z_o=100 \:mm$ the circle of confusion $CoC=49 \mu m$, $z_o^-=380 \: mm$ and $z_o^+=481 \: mm$. 

\begin{figure}
  \includegraphics{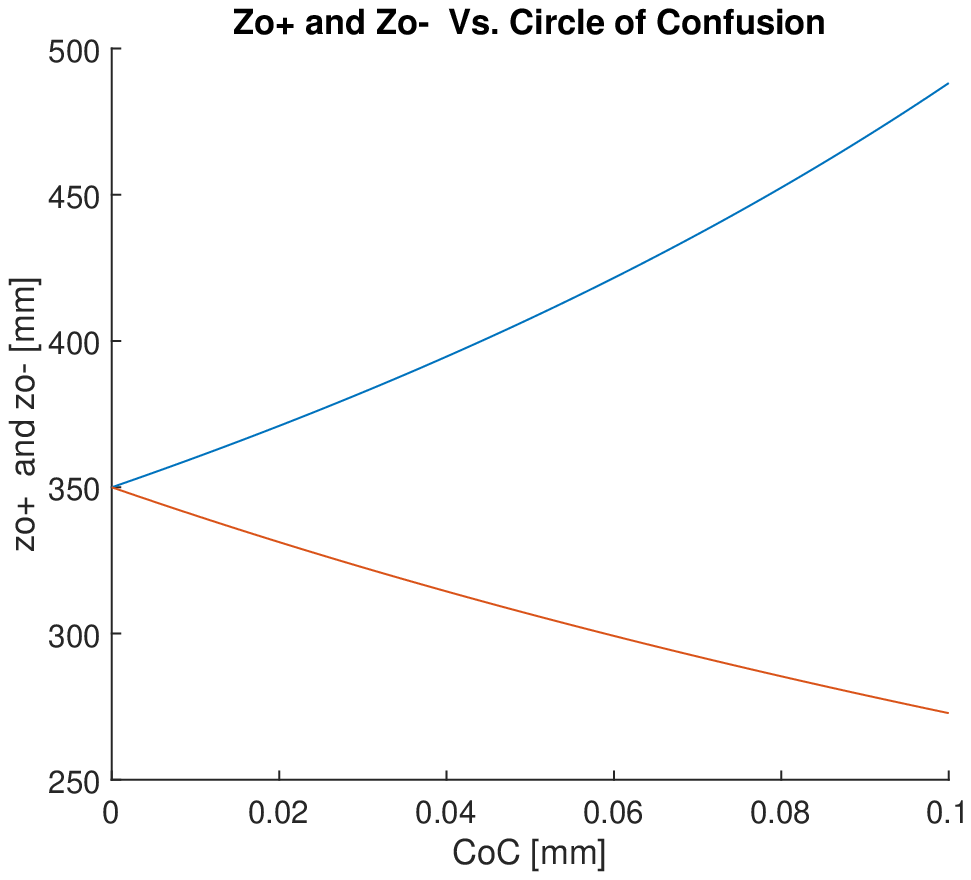}
 figure caption is below the figure
\caption{$z_o^-$ and $z_o^+ $ Vs. the circle of confusion where $z_o=424 \:mm, f=35 \: mm$ and $D_a=3.18 \:mm$.}
\label{fig:4}       
\end{figure}

\textit{The required laser repetition}: $f_{rep}=\frac{1}{0.00025}=4kHz$. 

\textit{The required pulse duration}: since the ratio of the depth of field to the working distance is not negligible, we use Eq.6 to find that the pulse width $\tau< \frac{13.5 \times 10^{-6}}{2 \cdot 0.1 \times 30}=6 \: \mu s$. We choose a laser with $\tau=0.01 \: \mu s$.

\textit{The required pulse energy}: We assume an elliptic Gaussian laser beam with $\omega_a=5 \: cm, \omega_b=15 \:cm$ that approximately fills the volume of interest $V$. The differential Mie scattering factor at $90 \deg$ was calculated by using Philip Laven free-ware program (www.philiplaven.com). The program  gives a Mie factor $ 10^{-16} \frac{W}{cm^2}$ at 1 meter for $d=15 \: \mu m$ and $\lambda=532 \: nm$. We recall that we need to calculate the signal level for a particle at the far edge of the depth of field (at $z_o^+$). The solid angle of this particle with the small aperture is calculated to be $\Omega_{par}=2.7 \cdot 10^{-7} \: str$. The image diameter of this particle on the image plane is $dM^++CoC=50 \: \mu m$. Therefore, the $FFR=0.1$. Plugging these numbers and those from table 2 into Eq.11, we obtain: 
$i_S=0.96 \times \frac{E_p}{2.3 \cdot 10^-5} \times 10^{-12} \times 2.7 \cdot 10^{-6} \times 0.8 \times 0.1 \times 0.23$
or $i_S= 6.8 \cdot 10^{-18} \times E_p$ Amperes. 
The electron count from a pixel for $E_p=1 \: mJ$ is found by dividing $i_S$ by the electron charge $e=1.6 \times 10^{-19}$. We obtain that we will have 0.042 electrons counts in one pixel (the image falls into several pixels) for a particle that is located at the far end of the depth of field. The electron noise count in the phantom camera is 30 electron. It means we would not be able to see a signal.  We would need a massive laser pulse (and average power). Alternatively, one will need to use techniques to re use a laser beam with a smaller cross section by multiple reflections, or use array of LEDs

\textit{Optical resolution and density}: The corrected Rayleigh distance at $z_o^+$, which considers the effect of the circle of confusion, is given by Eq.15. To calculate it, first we need to calculate the Rayleigh angle for two particles at the object plane : $\Delta \theta_R=1.22\frac{0.532 \cdot 10^-6}{0.00318} =204 \cdot 10^{-6} \: rad $. The distance between two particles at the $z_o^+$ plane is $D_R^+=\Delta \theta_R z_o^+=86 \: \mu m$.  The Rayleigh angle, corrected to the effect of the circle of confusion is given by Eq.14: $\theta_{R-CoC}^+=205.4 \cdot 10^{-6} \: rad$. The corrected Rayleigh distance is then $D_{R-CoC}^+= \theta_{R-CoC} z_o^+=98 \: \mu m$. 
  
To find the maximal particle density possible for our experiment, we first find the maximal particle density for a laser sheet with thickness $D_{R-CoC}^+$.  It is given by $C_{2D-PTV}=\frac{1}{\left( D_{R-CoC}^+\right)^3}$.  
The maximal density in cubic $cm$ is obtained by using Eq.13 with $D_R^o$ replaced by $D_{R-CoC}^+$:

\begin{equation}
C_{3D-PTV}=\frac{1}{\left( D_{R-CoC}^+ \right)^2} {z_o}=\frac{1}{\left(98 \cdot 10^-4\right)^2 \times 10}=1040 \: \frac{1}{cm^3}
\end{equation}
  
Due to the large depth of field, a point particle at $z_o^+$ has a large circle of confusion which compromises the signal level, the image resolution and as a consequence the spatial resolution of such a measurement.

%
%

\bibliography{Reference.bib}{}

%
%

\end{document}